\documentstyle[preprint,aps,psfig]{revtex}
\newcommand{\beq}{\begin{equation}}
\newcommand{\eeq}{\end{equation}} 
\newcommand{\beqa}{\begin{eqnarray}}
\newcommand{\eeqa}{\end{eqnarray}}
\newcommand{\ba}{\begin{array}}
\newcommand{\ea}{\end{array}}

\begin{document}
\draft

\widetext 
\title{Surface Effects in the Unitary Fermi Gas}
\author{L. Salasnich, F. Ancilotto, and F. Toigo} 
\address{Dipartimento di Fisica ``Galileo Galilei'' and CNISM, \\
Universit\`a di Padova, Via Marzolo 8, 35122 Padova, Italy} 

\maketitle

\begin{abstract} 
We study the extended Thomas-Fermi (ETF) 
density functional of the superfluid unitary Fermi gas. 
This functional includes a gradient term which is 
essential to describe accurately the surface effects of 
the system, in particular with a small number of atoms, 
where the Thomas-Fermi (local density) approximation fails. 
We find that our ETF functional gives density profiles 
which are in good agreement with recent Monte Carlo results 
and also with a more sophisticated superfluid density functional 
based on Bogoliubov-de Gennes equations.  
In addition, by using extended hydrodynamics equations of 
superfluids, we calculate the frequencies of collective surface oscillations 
of the unitary Fermi gas, showing that quadrupole and octupole modes 
strongly depend on the number of trapped atoms.
\end{abstract} 

\narrowtext 

\newpage 

\section{Introduction} 

In a system of spin $1/2$ interacting fermions, the unitary regime 
is commonly referred to as the situation in which the s-wave scattering 
length $a$ greatly exceeds the average interparticle separation, 
thus $n|a|^3\gg 1$, where $n$ is the number density \cite{giorgi}. 
In 2002 it was shown experimentally that such systems are 
(meta)stable \cite{hara}, and they have been studied extensively 
ever since \cite{giorgi}. 
On the basis of very accurate Monte Carlo results, we have 
recently introduced a simple but reliable 
extended Thomas-Fermi (ETF) density functional for the unitary Fermi gas 
at zero temperature \cite{etf}. 

In this paper we show new results obtained with our ETF functional. 
In particular, we analyze 
the density profiles of the balanced zero-temperature 
unitary Fermi gas confined by a harmonic trap, 
finding that our ETF functional gives density profiles 
which are in good agreement with recent Monte Carlo results 
\cite{doerte1,doerte2} and with a more sophisticated 
superfluid density functional based on Bogoliubov-de 
Gennes equations \cite{bulgac}. 
We then obtain extended hydrodynamics equations of 
superfluids by considering small, time-dependent 
perturbations of the ETF functional around its ground-state. 
We show that, under isotropic harmonic confinement, the quadrupole and 
octupole frequencies of a Fermi gas at unitarity 
depend on the number of confined atoms, contrary to the 
monopole and dipole frequencies, which are solely 
determined by the external potential. 

\section{Extended Thomas-Fermi density functional} 

At zero temperature the Thomas-Fermi (TF) energy functional\cite{giorgi} of 
a dilute and ultracold unpolarized two-component Fermi gas trapped 
by an external potential $U({\bf r})$ is 
\beq 
E_{TF} = \int  d^3{\bf r} \ 
n({\bf r}) \Big[  {\varepsilon}(n({\bf r});{ a_F}) 
+ U({\bf r}) \Big]  \ , 
\label{e-lda}
\eeq 
with ${\varepsilon}(n;a_F)$ bulk energy per particle, 
$n({\bf r})=n_{\uparrow}({\bf r}) + n_{\downarrow}({\bf r})$ total density 
($n_{\uparrow}({\bf r}) = n_{\downarrow}({\bf r})$) 
and $a_F$ the s-wave scattering length. The total number of fermions is 
\beq
N = \int d^3{\bf r} \ n({\bf r})   \; . 
\label{norma}
\eeq
By minimizing $E_{TF}$ one finds 
\beq
\mu(n({\bf r});{ a_F}) + U({\bf r}) = \bar{\mu} \; , 
\label{chem-lda}
\eeq 
with $\mu(n;{ a_F})=
{\partial (n {\varepsilon}(n;{ a_F}))\over \partial n}$ 
bulk chemical potential of a uniform system and $\bar{\mu}$ 
chemical potential of the non uniform system. 

For the uniform unitary Fermi gas\cite{bertsch} 
the s-wave scattering length $a_F$ diverges: 
\beq 
a_F \to\pm\infty \; ,  
\eeq    
and the only length characterizing the uniform system is the 
average distance between particles $n^{-1/3}$.  In this case:  
\beq 
{\varepsilon}(n;{ \xi}) = \xi 
{3\over 5} {\hbar^2 \over 2m} (3\pi^2)^{2/3} n^{2/3} = { \xi} \ 
{3\over 5} \ \epsilon_F \; , 
\label{eos} 
\eeq 
with $\epsilon_F$ Fermi energy of the ideal gas 
and ${ \xi}$ a universal parameter. 
The bulk chemical potential associated to Eq. (\ref{eos}) is 
\beq 
\mu(n;{ \xi}) = {\partial (n {\varepsilon}(n))\over \partial n} = 
{ \xi} {\hbar^2 \over 2m} (3\pi^2)^{2/3} n^{2/3} 
= { \xi} \ \epsilon_F \; . 
\label{mu}
\eeq

The TF functional must be extended to take into 
account other characteristic lengths related to the spatial variations 
of the density, besides the average particle separation. As a consequence, 
the energy per particle must contain additional terms, which scale as the 
square of the inverse of  these various lengths. For this reason, as a 
simple approximation, we add to the energy per particle the term 
\beq 
\lambda {\hbar^2 \over 8 m} {(\nabla n)^2\over n^2}  
= \lambda {\hbar^2\over 2m} {(\nabla \sqrt{n})^2 \over n}  \; . 
\label{grad-term} 
\eeq 
Historically, this term was introduced  by von Weizs\"acker \cite{von} 
to treat surface effects in nuclei. 
Moreover, according to the Kirzhnits expansion of the quantum kinetic operator 
in powers of $\hbar$ \cite{kirzhnits}, $\lambda$ must take the value 
$\lambda=1/9$ \cite{kirzhnits,sala-gradient}
for an ideal, noninteracting, Fermi gas. 
Here we consider ${ \lambda}$ as a {phenomenological parameter} 
accounting for the increase of kinetic energy due 
the spatial variation of the density. We also observe that 
the TF functional has a pathological behavior 
at the surface. In fact, the TF 
density profile becomes zero at a finite distance $r_{TF}$ 
from the center of the cloud, while according to quantum mechanics 
the density must goes to zero only at infinity. 
This pathology of the TF approximation is efficiently cured by 
the inclusion of the gradient term (\ref{grad-term}).

Other recent density-functional methods for unitary Fermi gas have been 
proposed in the last few years: the Bogoliubov-de Gennes (BdG) 
superfluid local-density approximation (SLDA) of Bulgac \cite{bulgac} 
and the Kohn-Sham (KS) density functional approach 
of Papenbrock \cite{papen}.   
We wish to point out that both the energy functionals 
proposed by Bulgac \cite{bulgac} and Papenbrock \cite{papen} 
are functionals of the density through single particle orbitals
(the BdG or KS orbitals). Therefore they can be used in actual numerical 
calculations only when the number of fermions is small, since 
they require a self consistent calculation of single-particle states 
whose number increases rapidly with the number of particles 
\cite{lipparini,ring}. For completeness, we remark that 
another density functional for the unitary Fermi gas 
has been proposed on the basis of the Haldane-Wu statistics 
\cite{zyl,bhaduri}. 

Our energy functional, that is the extended Thomas-Fermi (ETF) 
functional of the unitary Fermi gas, reads 
\beq 
E = \int d^3{\bf r} \ n({\bf r}) \Big[ 
{ \lambda} {\hbar^2 \over 8 m} {(\nabla n({\bf r}))^2\over 
n({\bf r})^2} 
+ { \xi} {3\over 5} {\hbar^2 \over 2m} (3\pi^2)^{2/3} n({\bf r})^{2/3}
+ U({\bf r})  \Big] \  \; .  
\label{e-dft}
\eeq
We stress that one encounters no severe limitation in the number of particles 
which may be treated through ETF functionals, 
since in this case the functional 
depends only on a single function of the coordinate, 
i.e. the particle density. 
By minimizing the ETF energy functional one gets:
\beq
\Big[ -{ \lambda} {\hbar^2\over 2m} \nabla^2 
+ \xi {\hbar^2 \over 2m} (3\pi^2)^{2/3} n({\bf r})^{2/3} 
+ U({\bf r}) \Big] \sqrt{n({\bf r})} = 
\bar{\mu} \ \sqrt{n({\bf r})} \; .  
\label{chem-dft} 
\eeq 
This is a sort of stationary 3D nonlinear Schr\"odinger (3D NLS) equation. 

The constants ${ \xi}$ and ${ \lambda}$ should be universal i.e. 
independent on the confining potential $U({\bf r})$. 
The value of the coefficient $\lambda$ is debated. In the papers of 
Kim and Zubarev \cite{kim} 
and Manini and Salasnich \cite{manini05} the authors set $\lambda=1$ 
over the full BCS-BEC crossover. More recently we have suggested 
$\lambda=1/4$ \cite{sala-josephson,sala-new,recent}.  
Note that ${ \lambda} \simeq 0.25$ was the preliminary prediction 
at unitarity of effective field theory \cite{rupak1}. 
We stress, however, that the more recent and accurate prediction 
of effective field theory based on $\epsilon$-expansion 
is $\lambda\simeq 0.17$ \cite{rupak2}. 

\section{Finding the universal parameters of the ETF functional}

To determine ${ \xi}$ and ${ \lambda}$ 
we look for the values of the two parameters which 
lead to the {best fit} of the ground-state energies 
obtained by Monte Carlo data for the unitary Fermi gas 
in a spherical harmonic potential 
\beq 
U({\bf r}) = {1\over2} m \omega^2 r^2 \; . 
\label{harmonic}
\eeq 
We use the more recent and reliable Monte Carlo results 
with $N$ even (complete superfluidity): 
the fixed-node diffusion Monte Carlo (FNDMC) of 
von Stecher, Greene and Blume \cite{doerte1}.  
After a systematic analysis \cite{etf} we find 
$$
\xi=0.455 \quad\quad \mbox{and} \quad\quad \lambda=0.13 
$$
as the best fitting parameters in the unitary regime. 
The value $\xi=0.455$ coincides with that obtained 
by Perali, Pieri, and Strinati \cite{perali} 
with beyond-mean-field extended BCS theory.  
Fixing $\xi=0.44$ (the Monte Carlo result for a uniform 
Fermi gas of Carlson {\it et al.} \cite{carlson})  
we find instead $\lambda=0.18$ \cite{etf}. 

In Fig. 1 we plot the ground-state energy $E$ of  
the unitary Fermi gas as a function of the number $N$ of atoms under harmonic 
confinement of frequency $\omega$. In the figure we compare 
the FNDMC data with even $N$ \cite{doerte1} with different 
density functionals: ETF results with best fit 
($\xi=0.455$ and $\lambda=0.13$); 
ETF results obtained from $\varepsilon$-expansion 
\cite{rupak2} ($\xi=0.475$ and $\lambda=0.17$); and also the 
TF (LDA) results ($\xi=0.44$ and $\lambda=0$). 
The figure clearly shows that the simple TF predictions 
are not fully reliable, while the best-fit ETF functional 
is extremely accurate.  

In Fig. 2 we report the comparison
between the density profiles obtained with our best-fit 
ETF functional and the FNDMC results \cite{doerte1,doerte2}. 
We have also included the SLDA results \cite{bulgac}.  
The figure shows that, near the surface, both ETF and SLDA methods 
are in very good agreement with the FNDMC predictions. 
Close to the center there are instead appreciable differences. 
Nevertheless, it is important to remind that in the FNDMC 
approach the statistical uncertainty is quite 
large at small distances from the center of the trap. 
Moreover, also SLDA values for central density depend quite 
a bit on the size of the BdG basis.  

For the sake of completeness, in Fig. 3 we plot the zoom 
of the density profile $n(r)$ near the surface for $N=20$, including also 
the TF prediction (dot-dashed line). The figure clearly shows 
that the TF data are not reliable near the surface. 

In our determination of $\xi$ and $\lambda$ 
we have analyzed the unitary gas with an even number $N$ of particles. 
Monte Carlo calculations show a clear odd-even effect 
(zig-zag effect): the ground state energy of $N$ odd particles 
in the isotropic harmonic trap is 
\beq 
E_N = {1\over 2} (E_{N-1} + E_{N+1}) + \Delta E_N   \; , 
\eeq
where the splitting $\Delta E_N$ is always positive. 
Dam Thanh Son has suggested \cite{son} 
that, given the superfluid cloud of even particles, 
the extra particle is localized where the energy gap is smallest, 
which is near the edge of the cloud. 
Dam Thanh Son has also found \cite{son} that, for fermions at unitarity, 
confined by a harmonic potential with frequency $\omega$, 
the odd-even splitting grows as 
\beq 
\Delta E_N = { \gamma} \ N^{1/9} \ \hbar \omega \; , 
\label{splitting}
\eeq 
where ${ \gamma}$ is an unknown dimensionless constant. 
After a systematic investigation of the FNDMC data we find that 
$\gamma =0.856$ gives the best fit \cite{etf}. 

\section{Extended superfluid hydrodynamics}

Let us now analyze the effect of the gradient term (\ref{grad-term})
on the dynamics of the superfluid unitary Fermi gas. 
At zero temperature 
the low-energy collective dynamics of this fermionic gas 
can be described by the equations of extended\cite{tosi,zaremba} 
irrotational and inviscid hydrodynamics:
\beqa 
{\partial n\over \partial t} &+& \nabla \cdot (n {\bf v}) = 0 \; , 
\label{hy1}
\\ 
m {\partial \over \partial t} {\bf v} &+& \nabla 
\Big[ - { \lambda} {\hbar^2\over 2m} {\nabla^2 \sqrt{n}\over \sqrt{n}} + 
\mu(n;{ \xi}) + U({\bf r}) + {m\over 2} v^2 \Big] = 0 \; .   
\label{hy2} 
\eeqa
They are the simplest extension of the equations of superfluid 
hydrodynamics of fermions\cite{giorgi} where ${ \lambda}=0$. 

The extended hydrodynamics equations can be written 
in terms of a {superfluid nonlinear Schr\"odinger equation} (NLSE), 
which is Galilei-invariant \cite{etf}. 
In fact, by introducing the complex wave function 
\beq 
\psi({\bf r},t) = \sqrt{n({\bf r},t)}\ e^{i\theta({\bf r},t)} \; , 
\label{wave}
\eeq
which is normalized to the total number $N$ of superfluid atoms, 
and taking into account the correct phase-velocity relationship 
\beq 
{\bf v}({\bf r},t) = {\hbar \over 2m} \nabla \theta({\bf r},t) \; , 
\label{velocity}
\eeq 
where $\theta({\bf r},t)$ is the phase of the condensate wavefunction 
of Cooper pairs, the equation 
\beq
i \hbar {\partial \over \partial t} \psi = 
\Big[ -{\hbar^2 \over 4 m} \nabla^2 + 2 U({\bf r}) + 
2 \mu(|\psi|^2;{ \xi}) 
+(1 - 4 { \lambda} ){\hbar^2\over 4m} 
{\nabla^2 |\psi|\over |\psi|} \Big] \psi \; , 
\label{nlse}
\eeq
is strictly equivalent to the equations of extended hydrodynamics. 

The extended hydrodynamics equations are the Euler-Lagrange equation 
of the following Lagrangian density 
\beq 
{\cal L} = - n \Big( {\dot{\theta}}  
+ {\hbar^2\over 8m} (\nabla \theta)^2 + U({\bf r}) + 
{\varepsilon}(n;{ \xi}) 
+ \lambda {\hbar^2 \over 8 m} {(\nabla n)^2\over n^2} \Big) 
\; ,  
\label{popov}
\eeq
which depends on the total number density $n({\bf r},t)$ and 
the phase $\theta({\bf r},t)$ \cite{etf}. 
In the case ${ \lambda}=0$ 
it is called the Popov Lagrangian of superfluid hydrodynamics \cite{popov}. 
 
Using, as previously, Eq. (\ref{wave}) we find that 
the extended Popov Lagrangian (\ref{popov}) 
is equivalent to the following one: 
\beq
{\cal L} = \psi^* 
\left(i {\hbar\over 2} {\partial \over \partial t} + {\hbar^2 \over 8m} 
\nabla^2 - U({\bf r}) 
- \varepsilon( |\psi|^2;{ \xi}) \right) \psi 
+ (1- 4{ \lambda}) 
{\hbar^2 \over 8 m} (\nabla |\psi|)^2   \; .  
\eeq 

\section{Sound velocity and collective modes} 

From the equations of superfluid hydrodynamics one finds the dispersion 
relation of low-energy collective modes of the 
uniform ($U({\bf r})=0$) unitary Fermi gas in the form 
\beq
{\Omega \over  q} = \sqrt{{ \xi}\over 3} v_F  \; , 
\eeq 
where $\Omega$ is the collective frequency, $q$ is the wave number and  
\beq 
v_F = \sqrt{2\epsilon_F\over m} 
\eeq
is the Fermi velocity of a noninteracting Fermi gas. 

The equations of extended superfluid hydrodynamics (or the 
superfluid NLSE) give also a correcting term \cite{etf}, i.e. 
\beq
 {\Omega \over  q} = \sqrt{{ \xi}\over 3} v_F 
\sqrt {1+ {3{ \lambda} 
\over { \xi}} 
\big( {\hbar q \over { 2 m v_F}}\big)^2 } \; ,   
\label{sound}
\eeq 
which depends on the ratio ${ \lambda}/{ \xi}$. 

In the case of isotropic harmonic confinement, Eq. (\ref{harmonic}),  
we study numerically the collective modes of the unitary Fermi gas 
by increasing the number $N$ of atoms. 
By solving the superfluid NLSE we find that 
the frequency $\Omega_0$ of the monopole mode ($l=0$) 
and the frequency $\Omega_1$ dipole mode ($l=1$) do not depend on $N$: 
$\Omega_0 = 2 \omega$ and $\Omega_1 = \omega$, 
as predicted by Castin \cite{castin}. 

We find instead that the frequencies $\Omega_2$ and $\Omega_3$ 
of quadrupole ($l=2$) and octupole ($l=3$) modes 
depend on $N$ and on the choice of the gradient 
coefficient $\lambda$. The frequencies are computed by solving 
the time-dependent Schr\"odinger equation (\ref{nlse}). 
We consider the initial state 
\beq 
\psi({\bf r},t=0) = e^{i \epsilon Q_l} \psi_{gs}({\bf r}) \; , 
\eeq  
where $\psi_{gs}({\bf r})$ is the ground-state wave function, 
$\epsilon$ is a small parameter and $Q_l$ is the operator 
which excites the surface mode with angular momentum $l$ 
(and $m_l=0$) \cite{lipparini}. For the quadrupole mode 
one has 
\beq 
Q_2 = 2z^2-x^2-y^2 
\eeq
while for the octupole mode the operator is 
\beq 
Q_3 = z(2z^2-3x^2-3y^2)\; . 
\eeq
Both $Q_2$ and $Q_3$ are expressed in cartesian coordinates \cite{lipparini}. 
In this way, a reliable value for the mode frequency can be 
obtained by following the real-time evolution of the system 
for rather short times, corresponding to few periods of oscillations at most. 

In Fig. 4 we plot the 
quadrupole frequency $\Omega_2$ as a function of the number 
$N$ of atoms under harmonic confinement. We consider 
three values of the gradient coefficient ${ \lambda}$. 
For ${ \lambda}=0$ one finds $\Omega_2=\sqrt{2} \omega$ \cite{giorgi}. 
As shown by the figure, this is the asymptotic value 
one finds only for a very large $N$ with a non-zero $\lambda$. 
Thus, we predict that finite-number effects can be detected 
in the unitary Fermi gas by measuring the quadrupole mode $\Omega_2$. 
This finite-number effect is stronger for surface modes 
with a larger angular quantum number $l$. This is shown in Fig. 5, 
where we plot the octupole mode $\Omega_3$ as a function of $N$. 

\section{Conclusions}

We have introduced an extended Thomas-Fermi (ETF) functional 
for the trapped unitary Fermi gas. 
By fitting FNDMC calculations we have determined the universal parameters 
${ \xi}$ and ${ \lambda}$ of ETF functional. 
ETF functional can be used to study ground-state density profiles 
in a generic external potential $U({\bf r})$.
We have also introduced a time-dependent version of the ETF functional: 
the extended superfluid hydrodynamics (or superfluid NLSE).
The superfluid NLSE can be used to investigate 
collective modes also for a small number of atoms. 
Our predictions suggest that surface effects are 
indeed very important for the unitary Fermi gas 
and that the physical observables strongly depend on them. 
Next-future experiments with less than one hundred degenerate Fermi 
atoms \cite{selim} across a Feschbach resonance 
can surely test our theoretical results. 

\vskip 0.5cm 

\noindent
{\it Acknowledgements} 
The authors thank Doerte Blume for making available her FNDMC data. 
LS acknowledges Sadhan K. Adhikari, Doerte Blume, Aurel Bulgac,  
Yvan Castin, Selim Jochim, Thomas Schaefer, 
and Eugene Zaremba for useful comments. This work has been 
partially supported by CARIPARO Foundation.

\newpage 

\begin{figure}[t]
\centerline{\psfig{file=etf-eps.eps,height=4.5in}} 
{\bf Fig. 1}. Ground-state energy $E$ for 
the unitary Fermi gas of $N$ atoms under harmonic 
confinement of frequency $\omega$. Symbols: FNDMC data with even $N$ 
\cite{doerte1}; 
solid line: ETF results with best fit ($\xi=0.455$ and $\lambda=0.13$); 
dashed line: ETF results obtained from $\varepsilon$-expansion 
\cite{rupak2} ($\xi=0.475$ and $\lambda=0.17$); dot-dashed lines 
TF (LDA) results ($\xi=0.44$ and $\lambda=0$). 
Energy in units of $\hbar \omega$.
\end{figure}

\newpage

\begin{figure}[t]
\centerline{\psfig{file=prof-compare.eps,height=4.5in}} 
{\bf Fig. 2}. Unitary Fermi gas under harmonic confinement 
of frequency $\omega$. 
Density profiles $n(r)$ for $N$ fermions obtained 
with our ETF ({solid lines}), SLDA \cite{bulgac} ({dashed lines}), 
and FNDMC \cite{doerte2} ({filled circles}). 
Lengths in units of $a_H=\sqrt{\hbar/(m\omega)}$.
\end{figure}

\newpage

\begin{figure}[t]
\centerline{\psfig{file=prof-surf.eps,height=4.5in}} 
{\bf Fig. 3}. Unitary Fermi gas with $N=20$ atoms 
under harmonic confinement of frequency $\omega$. 
Density profile $n(r)$ near the surface obtained 
with our ETF ({solid line}), SLDA \cite{bulgac} ({dashed line}), 
TF (dot-dashed line), and FNDMC \cite{doerte2} ({filled circles}). 
Lengths in units of $a_H=\sqrt{\hbar/(m\omega)}$.
\end{figure} 

\newpage

\begin{figure}[t]
\centerline{\psfig{file=quadrupole.eps,height=4.5in}} 
{\bf Fig. 4}. Quadrupole frequency $\Omega_2$ of 
the unitary Fermi gas (${ \xi}=0.455$) 
with $N$ atoms under harmonic 
confinement of frequency $\omega$. We use three different 
values of the gradient coefficient ${ \lambda}$. 
Notice that for ${ \lambda}=0$ (TF limit) one finds 
$\Omega_2=\sqrt{2} \omega$.
\end{figure} 

\newpage

\begin{figure}[t]
\centerline{\psfig{file=octupole.eps,height=4.5in}} 
{\bf Fig. 5}. Octupole frequency $\Omega_3$ of 
the unitary Fermi gas (${ \xi}=0.455$) 
with $N$ atoms under harmonic 
confinement of frequency $\omega$. We use two different 
values of the gradient coefficient ${ \lambda}$. 
Notice that for ${ \lambda}=0$ (TF limit) one finds 
$\Omega_3=\sqrt{3} \omega$.
\end{figure} 

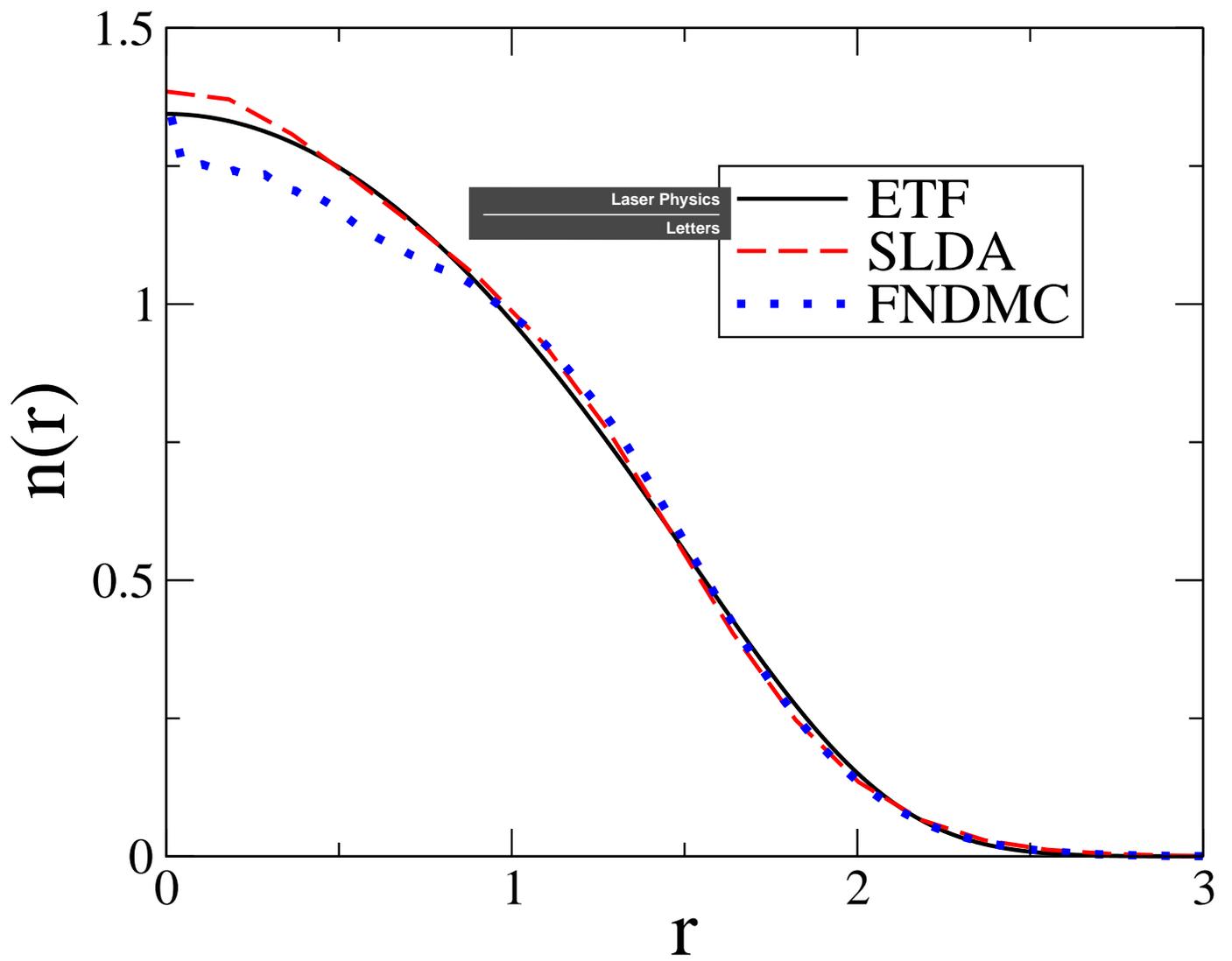
\end{document}